\documentclass[useAMS,usenatbib]{mn2e}

\usepackage{times,graphicx,amssymb,natbib}

\newcommand{\AIPS}{{$\cal AIPS\/$}}

\newcommand{\msol}{M_{\rm \odot}}

\title[A non-detection of XZ~Tau~C]{Triple trouble for XZ Tau: deep imaging with the Jansky Very Large Array}
\author[D. Forgan, R.J. Ivison, B. Sibthorpe, J.S. Greaves, E. Ibar]{D. Forgan$^{1}$\thanks{email:dhf@roe.ac.uk}, R.J. Ivison$^{1,2}$, B. Sibthorpe$^{3}$, J.S. Greaves$^{4}$, E. Ibar$^{5,6}$ \\
$^{1}$SUPA, Institute for Astronomy, University of Edinburgh, Blackford Hill, Edinburgh, EH9 3HJ, UK \\
$^{2}$European Southern Observatory, Karl-Schwarzschild-Strasse 2,
  D-85748 Garching, Germany\\
$^{3}$SRON Netherlands Institute for Space Research, Landleven 12,
  NL-9747 AD Groningen, the Netherlands \\
$^{4}$SUPA, School of Physics and Astronomy, University of St Andrews,
  North Haugh, St Andrews KY16 9SS, UK \\
$^{5}$Instituto de Astrof\'isica. Facultad de F\'isica. Pontificia Universidad Cat\'olica de Chile. Casilla 306, Santiago 22, Chile \\
$^{6}$Instituto de F\'isica y Astronom\'ia. Universidad de Valpara\'iso. Avda. Gran Breta\~na 1111. Valpara\'iso. Chile
}
\begin{document}

\date{Accepted}

\pagerange{\pageref{firstpage}--\pageref{lastpage}} \pubyear{}

\maketitle

\label{firstpage}

\begin{abstract} We present new observations of the XZ~Tau system made at high angular resolution (55\,milliarcsec) with the Karl G.\ Jansky Very Large Array (VLA) at a wavelength of 7\,mm.  Observations of XZ~Tau made with the VLA in 2004 appeared to show a triple-star system, with XZ~Tau~A resolved into two sources, XZ~Tau~A and XZ~Tau~C.  The angular separation of XZ~Tau~A and C (0.09\,arcsec) suggested a projected orbital separation of around 13\,AU with a possible orbital period of around 40\,yr.  Our follow-up observations were obtained approximately 8\,yr later, a fifth of this putative orbital period, and should therefore allow us to constrain the orbital parameters of XZ~Tau~C, and evaluate the possibility that a recent periastron passage of C coincided with the launch of extended optical outflows from XZ~Tau~A. Despite improved sensitivity and resolution, as compared with the 2004 observations, we find no evidence of XZ~Tau~C in our data.  Components A and B are detected with a signal-to-noise ratio greater than ten; their orbital motions are consistent with previous studies of the system, although the emission from XZ~Tau~A appears to be weaker.  Three possible interpretations are offered: either XZ~Tau~C is transiting XZ~Tau~A, which is broadly consistent with the periastron passage hypothesis, or the emission seen in 2004 was that of a transient, or XZ~Tau~C does not exist.  A fourth interpretation, that XZ~Tau~C was ejected from the system, is dismissed due to the lack of angular momentum redistribution in the orbits of XZ~Tau~A and XZ~Tau~B that would result from such an event.  Transients are rare but cannot be ruled out in a T~Tauri system known to exhibit variable behaviour.  Our observations are insufficient to distinguish between the remaining possibilities, at least not until we obtain further VLA observations at a sufficiently later time.  A further non-detection would allow us to reject the transit hypothesis, and the periastron passage of XZ~Tau~C as agent of XZ Tau A's outflows.  \end{abstract}

\begin{keywords}
methods: observational --- radio continuum: stars --- techniques: interferometric --- (stars:) binaries (including multiple): close
\end{keywords}

\section{Introduction}

XZ~Tau is a binary system composed of a T~Tauri star, XZ~Tau~A, with a cool companion, XZ~Tau~B, separated by approximately 0.3\,arcsec, at a distance of approximately 140\,pc from Earth \citep{Haas1990,Kenyon1994,Torres2009}.  Like many other T~Tauri stars, XZ~Tau~A drives collimated jets and optical outflows \citep{Mundt1990,Krist1997}.  {\it Hubble Space Telescope} imaging of these outflows shows nebular emission in the shape of an elongated bubble with expansion velocities of around 70\,$\mathrm{km\,s^{-1}}$ \citep{Krist1999}.  The substructure displayed by the bubble suggests its driver is episodic, with the cause attributed to a velocity pulse in the jet of XZ~Tau~A, triggered in the early 1980s \citep{Krist2008}.  These previous studies, particularly that of \citeauthor{Krist2008}, explored the possibility that the periastron passage of XZ~Tau~B could have caused the outflows (cf.\ \citealt{enc_outburst}).  However, this would require an eccentric orbit, which is inconsistent with observations of the A/B system, which instead suggest a circular, face-on orbit.  Also, the periastron passage of XZ~Tau~B would have occurred in the 1950s, too early to cause the outflow.

The periastron passage hypothesis was revived by more recent observations of the XZ~Tau system using the Very Large Array (VLA) by \citet{Carrasco-Gonzalez2009}.  Observations of the 7-mm continuum resolved XZ~Tau~A into two components, with the new component, XZ~Tau~C, separated by around 0.09\,arcsec (13\,AU).  The non-detection of XZ~Tau~C in the optical waveband was suggestive of a stellar object heavily embedded in a dusty envelope or disk.  While a single detection yielded no information on the orbit of component C around A, the existence of XZ~Tau~C increased the likelihood of a close approach to XZ~Tau~A, making it a potential trigger for the outflow.

\citet{Carrasco-Gonzalez2009} speculate that if the orbit of XZ~Tau~C is circular, and the total system mass is $\approx$1\,M$_\odot$, then the orbital period of XZ~Tau~C should be around 40\,yr. If the A/C system was close to apastron at the epoch of detection (2004), this would place XZ~Tau~C at periastron in the 1980s, as required, and a further ejection from XZ~Tau~A could be expected around 2020.  However, the data available to the authors was insufficient to confirm orbital parameters, and as such this explanation for the outflows was tentative.  Further observations of XZ~Tau~C at sufficiently later epochs would be required to either confirm or refute the periastron passage model for outflow generation.

To this end, we observed the XZ~Tau system at high angular resolution, using the newly upgraded Karl G.\ Jansky VLA at 7\,mm, to confirm the existence of XZ~Tau~C and constrain its orbital parameters.  The time interval between our observations (2012) and the previous observations (2004) corresponds to approximately one fifth of the potential orbital period of XZ~Tau~C.  These observations yield no detection of XZ~Tau~C, and the positions of XZ~Tau~A and B are consistent with the orbital solutions presented by previous studies.  This Letter is composed as follows: we describe the observations taken in \S\ref{sec:observations}; we discuss the results in \S\ref{sec:results}, and we summarise the work in \S\ref{sec:conclusions}.

\section{Observations}\label{sec:observations}

Our Q-band observations of the XZ~Tau system were obtained using the National Radio Astronomy Observatory's\footnote{This work is based on observations carried out with the VLA. The NRAO is a facility of the NSF operated under cooperative agreement by Associated Universities, Inc.} VLA , deployed in its most extended configuration (`A'), during three days in 2012 October (project code 12B-133, 2012 October 6, 11, 13, pointing centre R.A. 04 31 40.072, Dec +18 13 57.18). Each 1-hr observing block was scheduled during a period of excellent phase stability, with low wind speed. Dual-circular-polarisation data with a total bandwidth of 2\,GHz, comprising multiple 1-MHz channels centred at an observing frequency of 41\,GHz ($\approx$7\,mm), were recorded every 1\,s.

At the start of each observing block, the pointing accuracy of the antennas was refined using 3.6-cm continuum observations, just prior to observations of J0137+331 (3C\,48, used to calibrate the flux density scale) and J0431+2037 (used alongside J0431+1731 to track the complex gains on a timescale of 5\,min, and to correct for the bandpass response, and to test our likely astrometric accuracy). The positions of J0431+2037 and  J0431+1731 are known to $\approx$10 and $\approx$2\,milliarcsec, respectively, according to the NRAO Calibrator Manual.

Our data were edited using standard \AIPS\ procedures and calibrated using a recipe designed for high-frequency radio observations, as described in detail by \citet{Ivison2013}. The complex gains for our XZ~Tau scans were calibrated using both J0431+2037 and J0431+1731. A calibrated J0431+2037 dataset was also produced, using only J0431+1731.  Imaging of the calibrated $uv$ visibilities was also accomplished via \AIPS, using {\sc imagr}, with 10-milliarcsec pixels. The position of J0431+2037 was found to be $\alpha=04:31:03.76117\pm 0.00002$, $\delta=+20:37:34.2652\pm 0.0003$ (J2000), discrepant by 12 and 15\,milliarcsec in R.A.\ and Dec.\ from values in the NRAO Calibrator Manual, so consistent with the expected uncertainties. The astrometric measurement error here, for a signal-to-noise ratio of $\approx$55 and a synthesised beam with full width at half maximum (FWHM) of $\approx$55\,milliarcsec, is expected to be $\approx$1\,milliarcsec \citep{Ivison2007}, so we are dominated by systematics and our ability to transfer phase information accurately from J0431+1731.

The three individual XZ~Tau datasets produced self-consistent
images. Combining the data and employing a natural weighting scheme
yielded a $59\times 53$-milliarcsec (FWHM) synthesised beam, with a
north--south major axis at position angle $177^\circ$, which was used
to lightly clean the images; the resulting 1-$\sigma$ noise level (at
the pointing centre) was 23\,$\mu$Jy\,beam$^{-1}$. 

\section{Results \& Discussion}
\label{sec:results}

\begin{table*}
\caption{Detected Components in the XZ Tau System} 
\begin{tabular}{ccccc} 
\hline
Component & Identifier & R.A. & Dec. & Flux Density ($\mu$ Jy) \\  
\hline
N & XZ~Tau~B & 04 31 40.0811 $\pm$ 0.0002 & 18 13 56.890 $\pm$ 0.002 & 343.0$\pm$ 48.5 \\
S & XZ~Tau~A & 04 31 40.0953 $\pm$ 0.0001 & 18 13 56.712 $\pm$ 0.002 & 528.2 $\pm$ 55.5\\
\hline
\end{tabular}
\label{tab:components} 
\end{table*}

\begin{figure*} \begin{center}$ \begin{array}{cc} \includegraphics[scale=0.4]{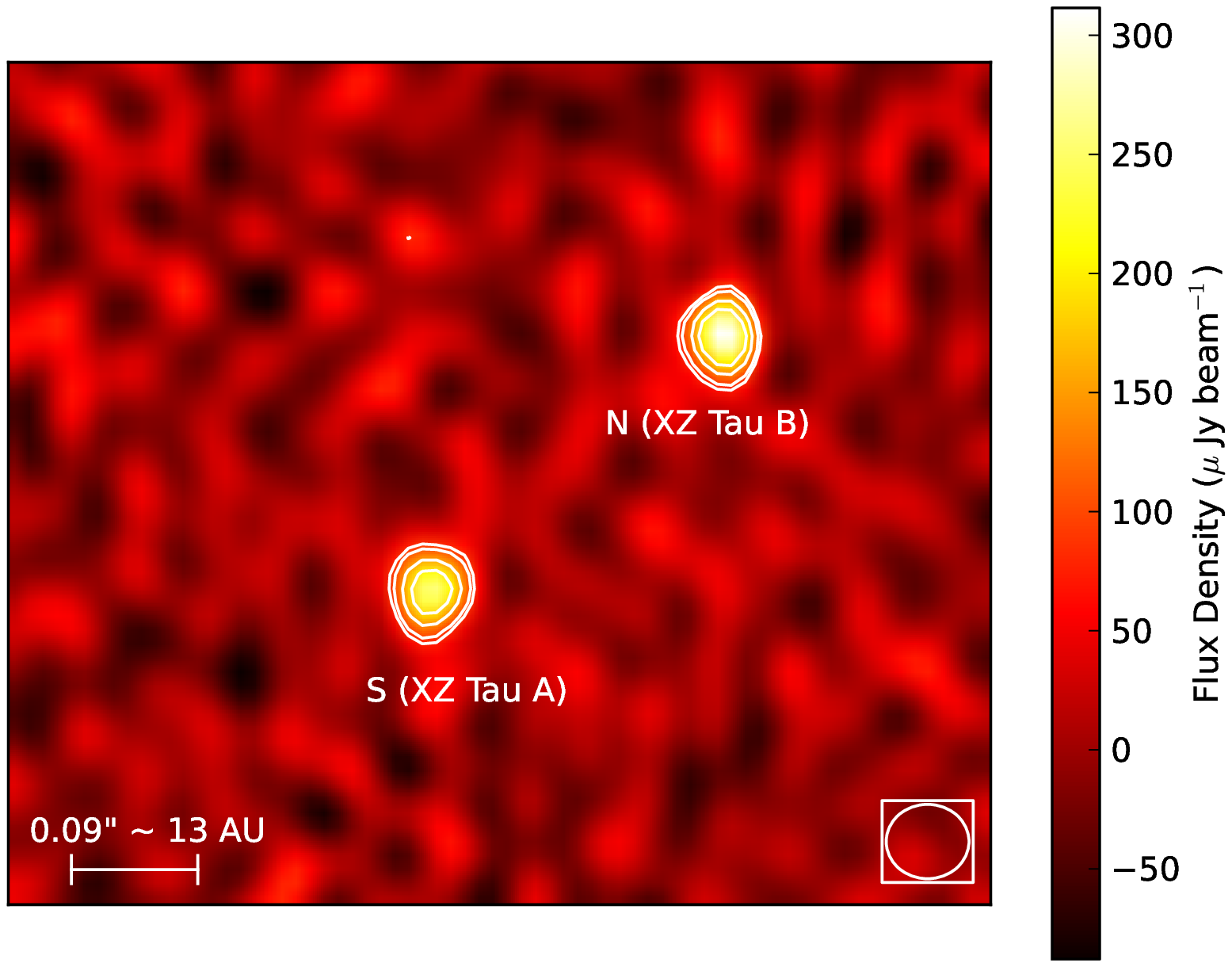} &
      \includegraphics[scale=0.4]{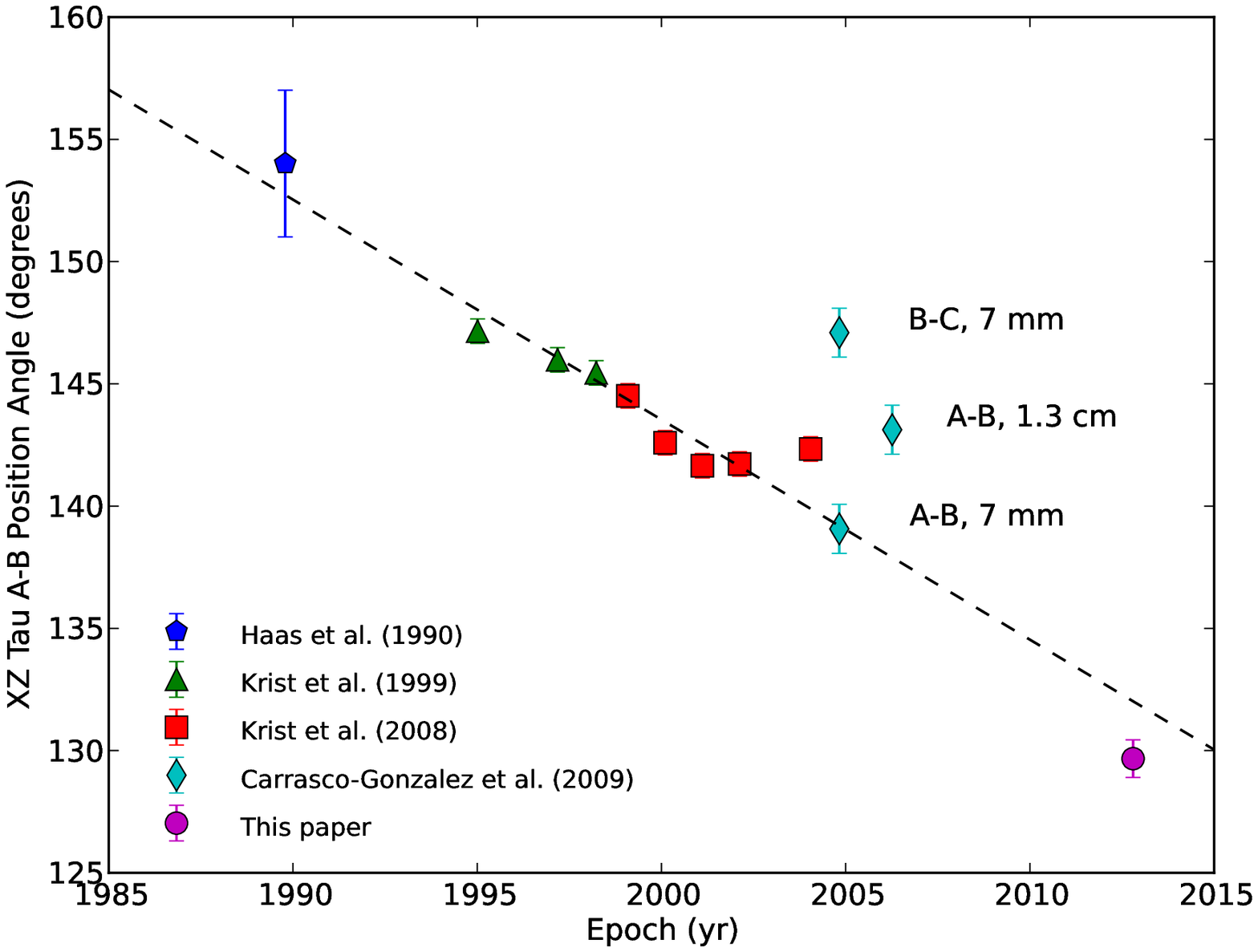} \\
      \includegraphics[scale=0.4]{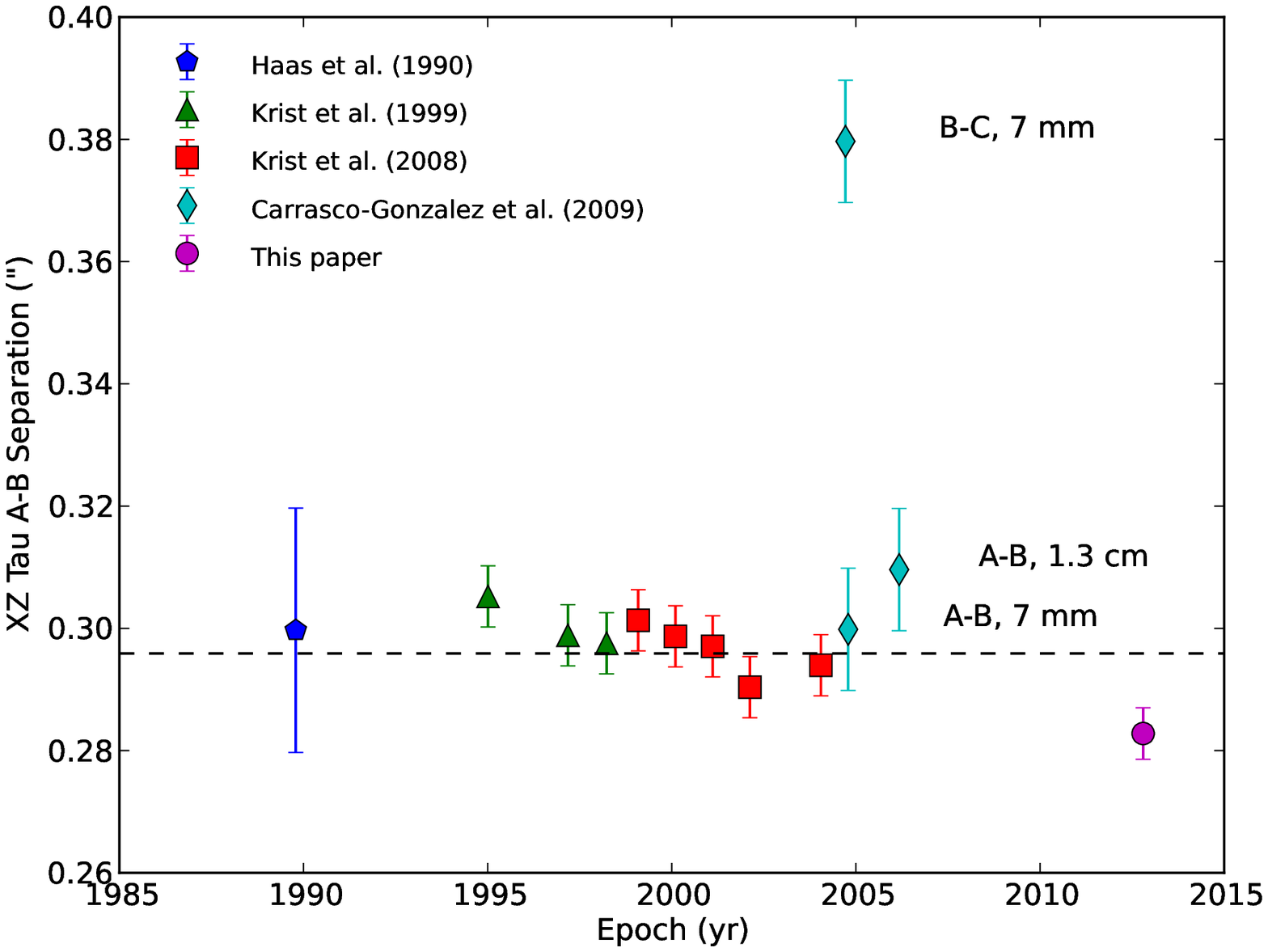} &
      \includegraphics[scale=0.4]{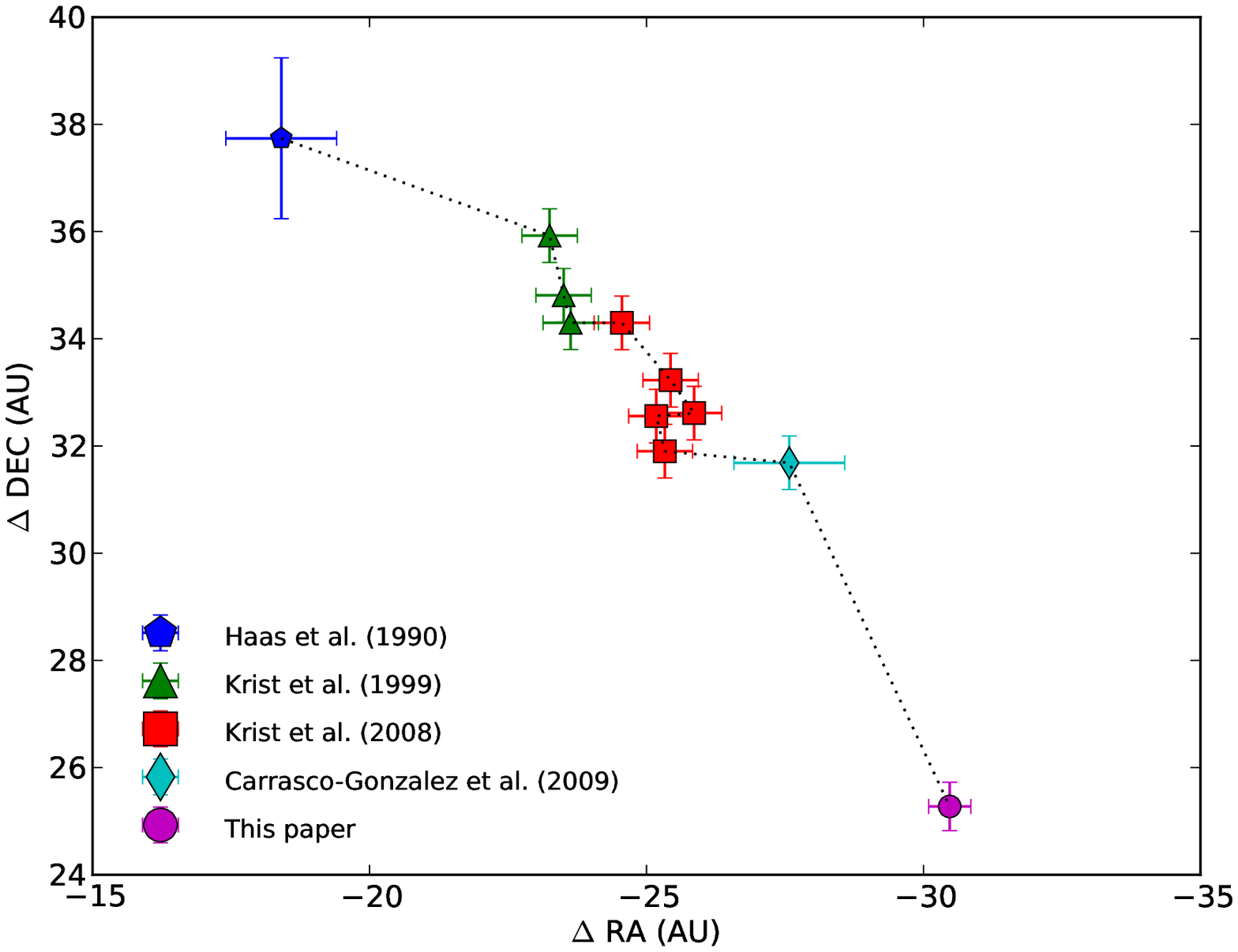} \\
    \end{array}$ \caption{Top left: XZ~Tau system, imaged at 7\,mm using VLA.  Contours are plotted for 80, 100, 150 and 200$\mu$Jy\,beam$^{-1}$.  The beam silhouette plotted refers to the synthesised beam described in section \ref{sec:observations}, with FWHM of $59\times 53$-milliarcsec.  Top right: change of position angle between XZ~Tau~A and B with epoch.  Also plotted is the separation between XZ~Tau~B and XZ~Tau~C as calculated by \citet{Carrasco-Gonzalez2009}. The dashed line corresponds to a best-fit orbital angular velocity of $-0.9$\,degrees\,$\mathrm{yr^{-1}}$.  Bottom left: projected angular separation of XZ~Tau~A and B with epoch.  The dashed line corresponds to a best-fit horizontal line of $0.296 \pm 0.002$\,arcsec.  Bottom right: evolution of the R.A.\ and Dec.\ of XZ~Tau~B relative to XZ~Tau~A, measured in the projected spatial distance at 140\,pc, as a function of time.  The dashed line corresponds to the best-fit orbit, assuming a circular, face-on configuration, with radius of 41.5\,AU.\label{fig:mainfig}}
  \end{center}
\end{figure*}

The top-left panel of Fig.~\ref{fig:mainfig} shows our 7-mm continuum map of the XZ~Tau system.  As it was our goal to detect the positions of all three components to high precision, our observations resolve out extended emission from XZ~Tau~A or B.  Table \ref{tab:components} shows the positions and flux densities of the two detected components, as derived from the 7-mm map.  Note that while XZ~Tau~B has a similar flux density as was recorded by \citet{Carrasco-Gonzalez2009}, XZ~Tau~A has a flux density less than half than previously, despite subsequent improvements to the sensitivity of the instrumentation used.

If the two detected components in this map correspond to XZ~Tau~A (southern component) and XZ~Tau~B (northern component), then their orbital parameters should be consistent with the literature.  We show the measured PA of the A--B binary in the top-right panel of Fig.~\ref{fig:mainfig}. Our PA of 129.67 degrees is consistent with the orbital angular velocity measured in \citet{Carrasco-Gonzalez2009} of $\Omega=-0.9 \pm 0.1$\,degrees\,$\mathrm{yr^{-1}}$, which we confirm by refitting the data with our extra point (and excluding the B--C data point from \citealt{Carrasco-Gonzalez2009}).

The separation of the components is also similar to previous measurements at 0.282\,arcsec.  If the orbit is assumed to be circular, then the best-fit line corresponds to a separation of $0.296 \pm 0.004$\,arcsec, or 41.4\,AU at 140\,pc.  As a sanity check, the relative R.A.\ and Dec.\ (bottom-right panel of Fig.~\ref{fig:mainfig}) can be fit with a circle of radius 41.5\,AU at 140\,pc, with an uncertainty in the fit of approximately 0.08\,AU or 0.5\,mas.  The fit does not improve significantly if the projected orbit is allowed to be elliptical, so we conclude that the orbit of XZ~Tau~A and B is close to face-on and circular.

These observations provide strong evidence that the two components detected in the map do correspond to XZ~Tau~A and B.  There is no detection of XZ~Tau~C, despite using the same facility as \citet{Carrasco-Gonzalez2009}, which has since been improved substantially; if XZ Tau C was present at the same flux density as previously recorded, our observations should have detected it with a signal-to-noise ratio of at least 3.   Subsequent reductions using the CASA pipeline also did not detect a third component.

There are four possible interpretations of the data:
\begin{enumerate}
\item  XZ~Tau~C was ejected from the system;
\item the emission seen by \citeauthor{Carrasco-Gonzalez2009} was due to a transient;
\item XZ~Tau~C is currently transiting (or being eclipsed by) XZ~Tau~A, and cannot be resolved;
\item XZ~Tau~C does not exist, and its previous detection was an artefact of the data calibration and analysis.
\end{enumerate}

\noindent We can quickly rule out the first possibility. No other sources that could correspond to an ejected XZ~Tau~C were detected.  For XZ Tau C to leave the field of view (at 7mm, this is approximately 1.1 arcsec) would require proper motions around 0.14 arcsec per year, corresponding to a speed of around 95 kilometres per second on the sky.  This is not an impossibly large velocity, but the orbits of XZ~Tau~A and B remain unperturbed, which is highly unlikely given the angular momentum redistribution an ejection would entail, as well as the typical recoil experienced by a binary when a third star is ejected \citep{Monaghan1976, Reipurth2000,Reipurth2012}.  

We cannot rule out emission from a transient, particularly not in a system known to exhibit variability. In general, young stellar objects display transient emission over the wavelength range probed by the VLA (see e.g.  \citealt{Dzib2013}).  Conversely, we note that extragalactic transient events are rare \citep[e.g.][]{Carilli2003, Frail2012, Mooley2013}.


For XZ~Tau~C to have triggered the outflows from XZ~Tau~A at the appropriate epoch, \citet{Carrasco-Gonzalez2009} assume that the orbit of XZ~Tau~C should be nearly circular ($e\approx 0.1$) and face-on, where they assume the total mass of A and C is approximately $1\msol$ and that XZ~Tau~C was at apastron during their observations (2004.8, with an apastron radius of approximately 13\,AU).  However, if the orbit is edge-on rather than face-on, an $e=0.1$ orbit with the same apastron radius is approximately consistent with XZ~Tau~C being either in transit or in eclipse.  If this interpretation is correct, then the periastron passage hypothesis may also be correct.  However, transits of A by C remain possible with other selections of orbital parameters, and cannot be ruled out thanks to the non-detection reported here.

On the other hand, this intepretation would require XZ~Tau~C to orbit almost perpendicular to the A--B binary plane, leaving it vulnerable to strong perturbations from the Kozai-Lidov mechanism \citep{Kozai1962,Lidov1962}, generating significant coupled oscillations in the star's eccentricity and inclination \citep{Naoz2013}.  Given that the oscillation timescale is several orders of magnitude larger than the interval between epochs of observation, this possibility cannot be ruled out.

This leaves us with the final interpretation -- that the previous detection of XZ~Tau~C was erroneous.  This is consistent with the non-detection of XZ~Tau~C at optical wavelengths.  While it is true that XZ~Tau~C could have been a highly embedded star, as \citet{Carrasco-Gonzalez2009} suggest, a non-detection at 7\,mm is not consistent with this interpretation. Observations at a sufficiently later epoch are required to decide which of our interpretations is most likely.

\section{Conclusions}
\label{sec:conclusions}

\noindent We report observations of the XZ~Tau system, recently observed to possess a third component, XZ~Tau~C \citep{Carrasco-Gonzalez2009}.  This third component had been proposed as a potential driver for the outflows generated by XZ~Tau~A, which could have been generated by a periastron passage of this new object in the 1980s.  Our new observations were made using the VLA in its most extended configuration, after a period of roughly one fifth of the object's assumed orbital period, and can thereby constrain the orbit of XZ~Tau~C and test the periastron passage theory.

However, our observations yield no detection of XZ~Tau~C.  XZ~Tau~A and B are both detected with a high degree of confidence, with positions and orbits consistent with those described in the literature.

Three potential interpretations of the data are possible.  Of these, the most prosaic is that XZ~Tau~C does not exist, and was an artefact of reducing interferometric data, which is consistent with its non-detection in the optical. Alternatively, XZ~Tau~C may have been caught whilst transiting XZ~Tau~A, which is consistent with the orbital requirements for XZ~Tau~C to trigger outflows at periastron passage.  This second interpretation would require the hierarchical triple to possess a large mutual inclination, and hence be dynamically unstable, but on timescales that are sufficiently long to remain consistent with the observations. Finally, we cannot rule out the possibility of a transient event local to this T~Tauri system, a system known to be variable. To distinguish between these potential interpretations, determining the nature of XZ~Tau~C, requires a very brief, future observation using the VLA in A~configuration.  If the transit hypothesis is true, future observations should be able to detect XZ~Tau~C once it has moved away from XZ~Tau~A.  If observations at this time fail to detect XZ~Tau~C, this would be sufficient to reject it as a cause of outflows being generated by XZ~Tau~A.  

\section*{Acknowledgments}

\noindent DF gratefully acknowledges support from STFC grant ST/J001422/1.  RJI acknowledges support in the form of ERC Advanced Investigator programme, {\sc cosmicism}.  EI acknowledges funding from CONICYT/FONDECYT postdoctoral project N$^\circ$:3130504.  The authors thank Claire Chandler for reductions made using the CASA pipeline.

\bibliographystyle{mn2e} 
\bibliography{no_XZTauC}

\appendix

\label{lastpage}

\end{document}